\documentstyle[prc,aps,epsf]{revtex}

\def\Journal#1#2#3#4{{#1} {\bf #2}, #3 (#4)}
\def\NCA{\em Nuovo Cimento}
\def\NPA{{\em Nucl. Phys.} A}

\def\PRC{{\em Phys. Rev.} C}

\begin{document}

\draft 

\title{ELECTRON SCATTERING WITH A \\SHORT-RANGE CORRELATION MODEL
\footnote{Talk presented by G. Co' at the ``4th. Workshop on
Electromagnetcally Induced Two-Hadron Emission'', Granada, Spain, 1999.}
}

\author{J.E. AMARO$^1$, F. ARIAS DE SAAVEDRA$^1$, A.M. LALLENA$^1$,
G. CO'$^2$, \\A. FABROCINI$^3$ and S. RASHAD$^{2,4}$ \\
$^1$Departamento de F\'{\i}sica Moderna, Universidad de Granada, Granada,
Spain. \\
$^2$Dipartimento di Fisica, Universit\`a di Lecce, and\\
INFN sez. Lecce, Lecce, Italy. \\
$^3$Dipartimento di Fisica, Universit\`a di  Pisa, and \\
INFN sez. Pisa,  Pisa, Italy. \\
$^4$Department of Physics, University of Assiut, Assiut, Egypt.
}

\maketitle

\begin{abstract}
 The inclusive electromagnetic responses in the
  quasi--elastic region are calculated with a model which considers
  the terms of the cluster expansion containing a single correlation
  line. The validity of this model is studied by comparing, in nuclear
  matter, its results with those of a complete calculation. Results
  in finite nuclei for  both one-- and two--nucleon
  emission are presented. 
\end{abstract}

\section{Introduction}
Aim of the nuclear many--body theories is the calculation of the
properties of finite nuclear systems by having as only input the bare
nucleon--nucleon interaction. 
The approach we follow to solve the nuclear many--body problem
is based upon the variational principle 
\begin{equation}
\label{efunc}
\delta E[\Psi_0]=\delta \frac {\langle\Psi_0 |H|\Psi_0\rangle} {\langle\Psi_0 |\Psi_0\rangle}
=0
\,\,.
\end{equation}
The solution of eq. (\ref{efunc}) is exactly that of
the Schr\"odinger equation if there are
no limitations on the Hilbert space spanned to search for the minimum
of the energy functional. 
In our approach we search for the minimum within a
subspace composed by states of the form
\begin{equation}
\label{ansa1}
|\Psi_0\rangle=F |\Phi_0\rangle \,\,\, ,
\end{equation}

\noindent
where $|\Phi_0\rangle$ is a Slater determinant, formed by a set of
orthonormal single particle wave functions, and $F$ is a many--body
correlation function. 

The quantities to be compared to the experiment are
calculated by evaluating the mean value of the corresponding operators
between the states of eq. (\ref{ansa1}) fixed at the minimum of the energy
functional: 
\begin{equation}
\label{ave}
 \langle Q \rangle=\frac {\langle\Phi_0 |F^+ Q F |\Phi_0\rangle} {\langle\Phi_0|F^+F |\Phi_0\rangle}
 \,\,\, ,
\end{equation}
where  we have indicated with
$Q$ a generic operator associated to an observable. 

A good variational ansatz for realistic calculations requires that
the correlation function $F$ has the 
same operatorial dependence of the Hamiltonian. 
In this report, however, we assume that the many--body correlation 
is a scalar
function which can be written as a product of two-body correlation
functions:
\begin{equation}
\label{cormany}
F({\bf r}_1, {\bf r}_2, .... ,{\bf r}_A,)=
\prod_{i<j}^{A}f({r_{ij})}
\end{equation}
where $r_{ij}$ is the distance between the nucleons $i$ and $j$.
The choice (\ref{cormany}) of the correlation function
leads to the following expression for the energy functional:
\begin{eqnarray}
\nonumber
E[F\Phi_0] &=&  \frac {\langle\Phi_0 |H \prod_{i<j}^A f^2(r_{ij})
|\Phi_0\rangle} 
{\langle\Phi_0|\prod_{i<j}^A f^2(r_{ij})|\Phi_0\rangle} \\
&=&
\label{ce}
\frac {\langle\Phi_0 |H [1+h(r_{12})] [1+h(r_{13})] .....
  [1+h(r_{23})]... |\Phi_0\rangle}  
{\langle\Phi_0 |[1+h(r_{12})] [1+h(r_{13})] .....
  [1+h(r_{23})]... |\Phi_0\rangle}  
\end{eqnarray}
In the above equation we have used the function 
$h(r_{ij})=f^2(r_{ij})-1$ 
to show the mechanism of the cluster expansion.

The simplest term is  
the mean value of the hamiltonian between two uncorrelated states, which
is obtained by picking up all the ``ones'' in the various terms in
eq. (\ref{ce}).
More complicated terms are obtained considering a single
$h(r_{ij})$ function multiplying the ``ones''. The procedure continues
considering terms with two, three, etc. $h(r_{ij})$ functions. We make
a classification of the various terms based upon their topology.
With the Fermi Hypernetted Chain (FHNC) technology we calculate all
the terms of a certain type.
Details of the application of the FHNC to the
description of the ground state of doubly closed shell nuclei can be
found in refs. \cite{co92}$-$\cite{fab98}. 
The recent progresses of
these calculations, done with state dependent correlations,
are presented in another contribution to this conference \cite{ari99}.

\section{The model: ground state}
\begin{figure}[ht]
\begin{center}
\hspace*{-.8cm}
\leavevmode
\epsfysize = 350pt
\epsfbox[70 200 500 650]{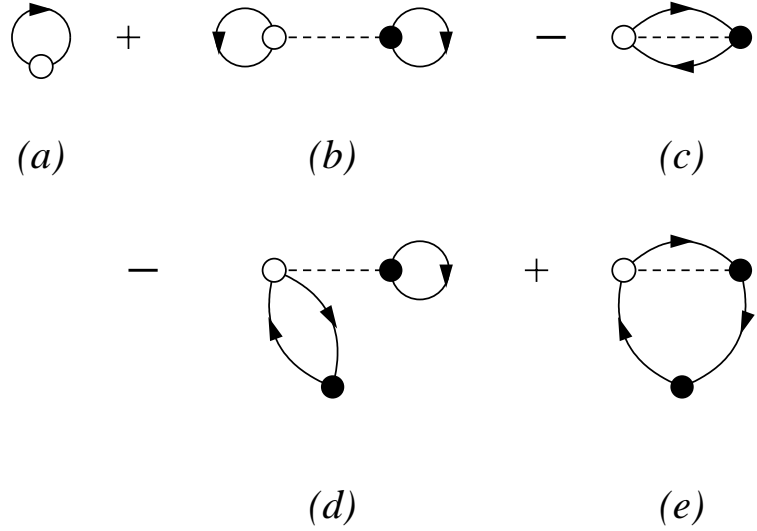}
\end{center}
\vspace{-5cm}
\caption[]{\small Diagrams representing the terms considered in our
  model for the calculation of the density distribution.  }
\label{diag1}
\end{figure}

In the theoretical framework we have presented, 
the evaluation of an operator mean value,
eq. (\ref{ave}), requires the use of the full FHNC computational
scheme. 
For the evaluation of the one--body density distribution, defined as 
\begin{equation}
\label{rhof}
\rho({\bf r}_1)= \frac{A}{\langle\Psi_0|\Psi_0\rangle}
\int d^3 r_2 ..... \int d^3 r_A
\Psi_0^*({\bf r}_1,....,{\bf r}_A)\Psi_0({\bf r}_1,....,{\bf r}_A)
\,\, ,
\end{equation}
we have developed a simplified model \cite{co95,ari97}.
After performing the cluster expansion of both numerator and denominator
to eliminate the so-called unlinked diagrams, 
we retain the terms with a single correlation function $h$.  
The expression for the density obtained in this approximation is:
\begin{eqnarray}
\label{rho1}
\nonumber
&&\rho_1({\bf r}_1)= 
\int d^3 r_2 ..... \int d^3 r_A \\
&&\Phi_0^*({\bf r}_1,....,{\bf r}_A)
\left[ 1 + \sum_{i=2}^A h(r_{1i}) + \sum_{i=2}^A \sum_{j>i}^A h(r_{ij})
\right]_L
\Phi_0({\bf r}_1,....,{\bf r}_A)
\end{eqnarray}
where the subindex $L$ indicates that only the linked diagrams 
have to be calculated. 
The diagrams representing the terms considered in our
model are shown in fig.~\ref{diag1}. In this figure the white circles
represent the point ${\bf r}_1$ which is not integrated, the black
circles the points  ${\bf r}_i$ and  ${\bf r}_j$ which are integrated
in eq. (\ref{rho1}), the dashed line the function $h(r_{ij})$ and the
oriented lines are related to the single particle wave functions.

The density calculated in our model is correctly normalized. This can
be seen considering that, since the uncorrelated term, corresponding
to the diagram $(a)$ of fig. \ref{diag1}, is already correctly
normalized, the contribution to the normalization
of the other terms should be zero. 
In the diagrammatic picture of fig.~\ref{diag1} the
integration on  ${\bf r}_1$ transform in black the white dot. Since in
the three point diagrams
$(d)$ and $(e)$ of the figure the white circles are
not reached by the correlation line $h(r_{ii})$, 
the integration implies that the single
particle wave functions connected to these points should be the
same. Therefore the contribution of the diagram
$(d)$ cancel that of the diagram $(b)$ 
as well as the contribution of the diagram $(e)$ that of the diagram
$(c)$. 

We checked the validity of our model by comparing
the density distributions we have obtained with those produced
by a FHNC calculation. 
The comparison has been done using correlations and single particle
wave functions given by the minimization of the energy functional
generated  by the semi-realistic Afnan and Tang
S3 interaction \cite{afn68}.

\begin{figure}[ht]
\begin{center}
\vspace*{-2.5cm}
\hspace*{.8cm}
\leavevmode
\epsfysize = 350pt
\epsfbox[70 200 500 650]{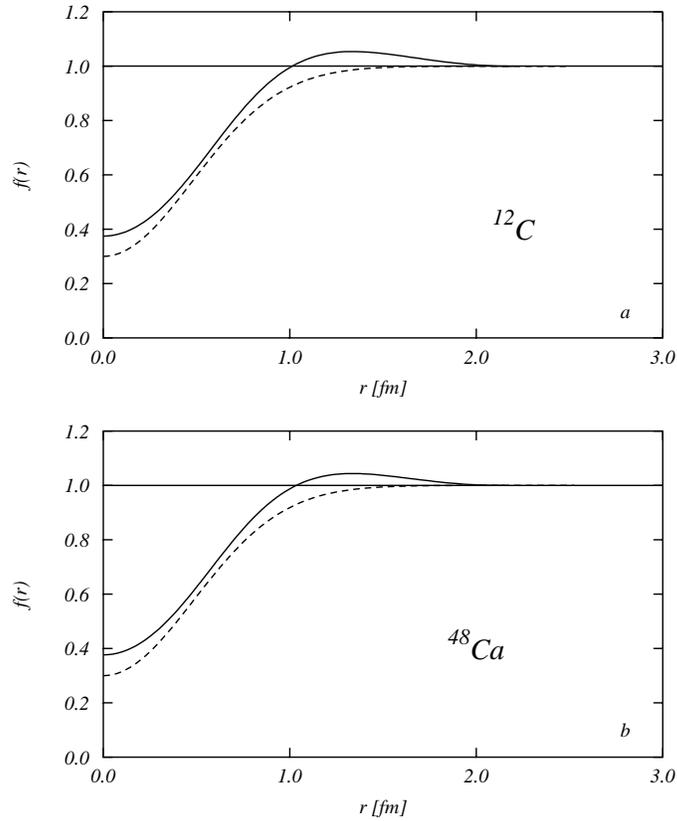}
\end{center}
\vspace{2cm}
\caption{\small Correlation functions obtained for $^{12}$C and $^{48}$Ca
  using the semi-realistic S3 nucleon--nucleon interaction. 
Dashed lines are gaussian correlations and full lines
  are the Euler correlation functions.
}
\label{corr1}
\end{figure}

In fig. \ref{corr1} the correlations functions
for the $^{12}$C and $^{48}$Ca nuclei are presented.
The dashed lines show the results of a gaussian parameterization
(two free parameters), while the full lines 
have been obtained with a procedure, we call it Euler
minimization, consisting in minimizing the energy
functional calculated up to the second order. 
In this last calculation there is only one free parameter: the healing
distance. 

\begin{figure}[ht]
\begin{center}
\vspace*{-6cm}
\hspace*{-.8cm}
\leavevmode
\epsfysize = 350pt
\epsfbox[70 200 500 650]{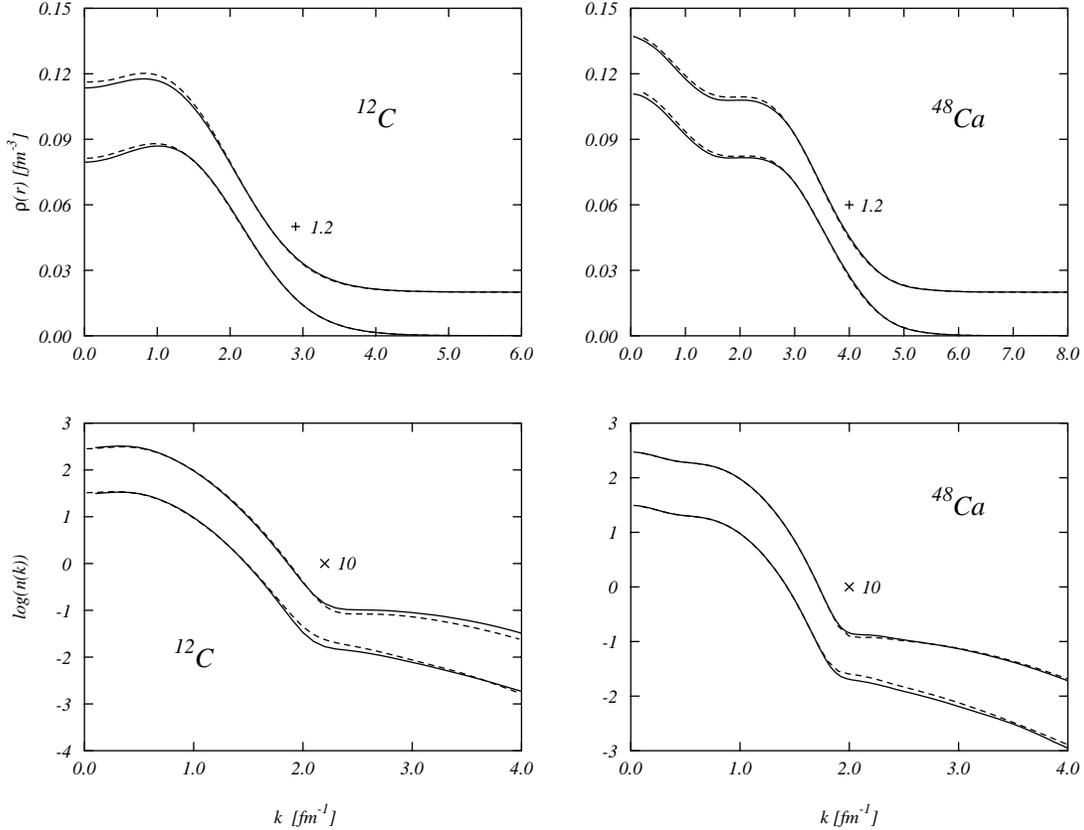}
\end{center}
\vspace{5cm}
\caption{\small Comparison between densities (upper panels) and momentum
  distributions (lower panels) calculated with our model (dashed lines)
  and with a FHNC procedure (full lines). In all the panels the upper
  curves show the calculations done with Euler correlations, and the
  lower curves those done with gaussian correlations. 
}
\label{dens}
\end{figure}

The comparison between the results of the FHNC calculation and those
of our model is done in fig. \ref{dens}. The 
one--body densities are compared in
the upper panels for calculations done with both kinds of
correlations. The agreement between the result of the two calculations
is very good. 
Since the one--body densities are the diagonal part of the density
matrix, in order to test our model 
also on the off diagonal part of this matrix, the comparison with the
full calculation has been done also for the momentum
distributions. These results are shown in the lower panels of
fig. \ref{dens}, and also in this case the agreement between the
two calculations is rather satisfactory.

%
%
%

%
%

%
%

\section{The model: excited states}

The description of nuclear responses in the quasi--elastic region
 is based upon the approach developed by Fantoni and Pandharipande
 \cite{fan87} and applied to nuclear matter. 
The basic ansatz consists in assuming that the nuclear excited states
 can be described as a product of a Slater determinant and a
 many--body correlation function which is the same  
 correlation function describing the ground state:
\begin{equation}
\label{ansa2}
|\Psi_f\rangle=F |\Phi_f\rangle \,\,\, .
\end{equation}
In eq. (\ref{ansa2}), $|\Phi_f\rangle$ indicates a Slater determinant 
which differs from  $|\Phi_0\rangle$ by the fact that
a certain number of hole single particle states have
been substituted with particle wave functions.

The many--body nuclear response at energy $\omega$ and momentum
transfer ${\bf q}$ to a an external field $Q({\bf q})$ can be written
as:
\begin{equation}
\label{resp}
R(\omega,{\bf q})=\sum_f 
\frac {\langle\Psi_0 |Q^+({\bf q})|\Psi_f\rangle\langle\Psi_f |Q({\bf q})|\Psi_0\rangle}
      {\langle\Psi_f |\Psi_f\rangle\langle\Psi_0 |\Psi_0\rangle}\,\,\, \delta(E_f-E_0-\omega)
\end{equation}
where the sum is understood to run on all the nuclear excited states. 
The evaluation of eq. (\ref{resp}) requires the calculation 
of the full cluster expansion \cite{fan87}.

Like in the case of the density distribution, instead of performing
the full expansion, after the elimination of the unlinked
diagrams, we consider only the terms containing a single correlation
function $h(r_{ij})$.  

In the case the nuclear final state is characterized by 
one particle above the Fermi surface, 
the matrix element to be calculated are of the kind:
\begin{equation}
\label{1p1h}
\xi_{ph}({\bf q})=\frac {\langle\Phi_{ph}|F^+ Q^+({\bf q}) F|\Phi_0\rangle} 
                        {\langle\Phi_{ph}|F^+ F|\Phi_0\rangle} 
\,\, ,
\end{equation}
where we have indicated with $|\Phi_{ph}\rangle$ the Slater determinant
where the hole single particle wave function $h$ has been substituted
by the particle  wave function $p$. 

\begin{figure}
\begin{center}
\vspace*{4.3cm}
\hspace*{-2.0 cm}
\leavevmode
\epsfysize = 300pt
\epsfbox[70 200 500 650]{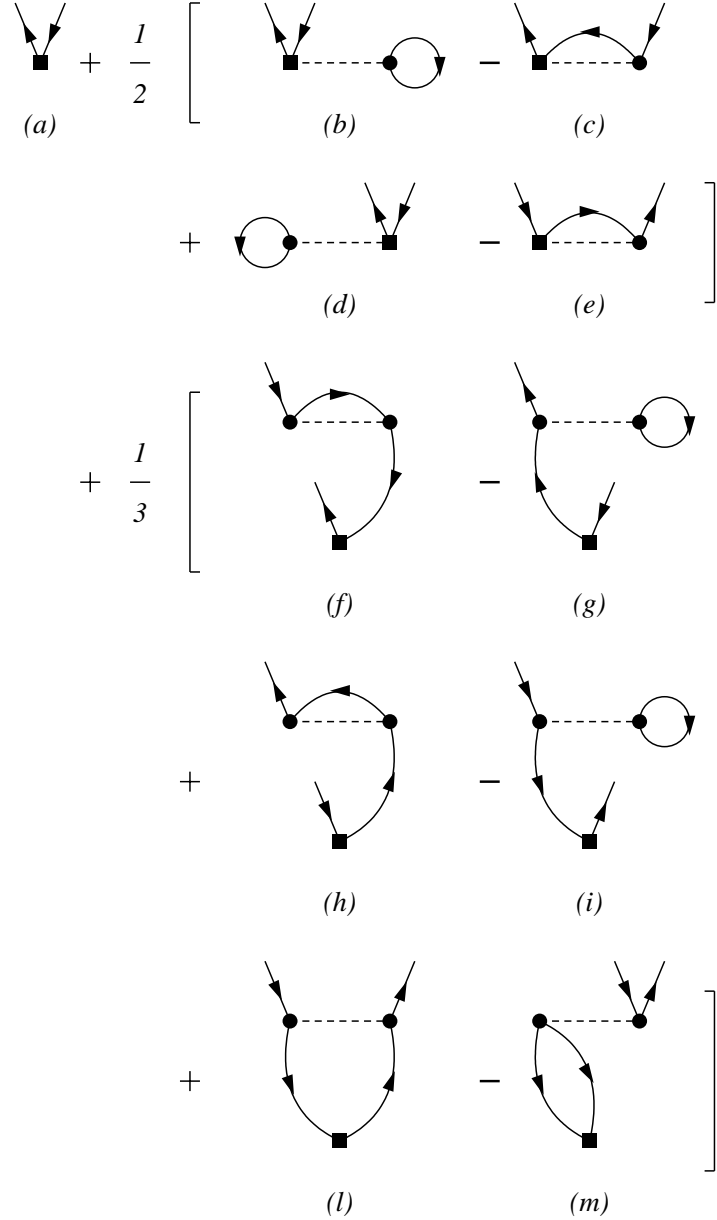}
\end{center}
\vspace{3.2cm}
\caption{\small Diagrams considered in the calculation of the one-particle
  one-hole response. The symbols have the same meaning of those in
  fig~\ref{diag1}. The black squares represent the point where the
  external field is acting on.  
}
\label{diag2}
\end{figure}

If the $Q$ operator is the charge operator, eq. (\ref{1p1h}) represents, 
in the limit for $q \rightarrow 0$ and $p \rightarrow h$,
the normalization condition of the density. 
This fact allows us to identify the diagrams which we should consider
to have a properly normalized many--body wave function.
The diagrams we consider are shown in fig. \ref{diag2}. One can
reconstruct the diagrams of fig. \ref{diag1} by closing the particle
and hole lines.

%
%
%

%
%
\begin{figure}[ht]
\begin{center}
\vspace*{0.5cm}
\hspace*{-1.5cm}
\leavevmode
\epsfysize = 300pt
\epsfbox[70 200 500 650]{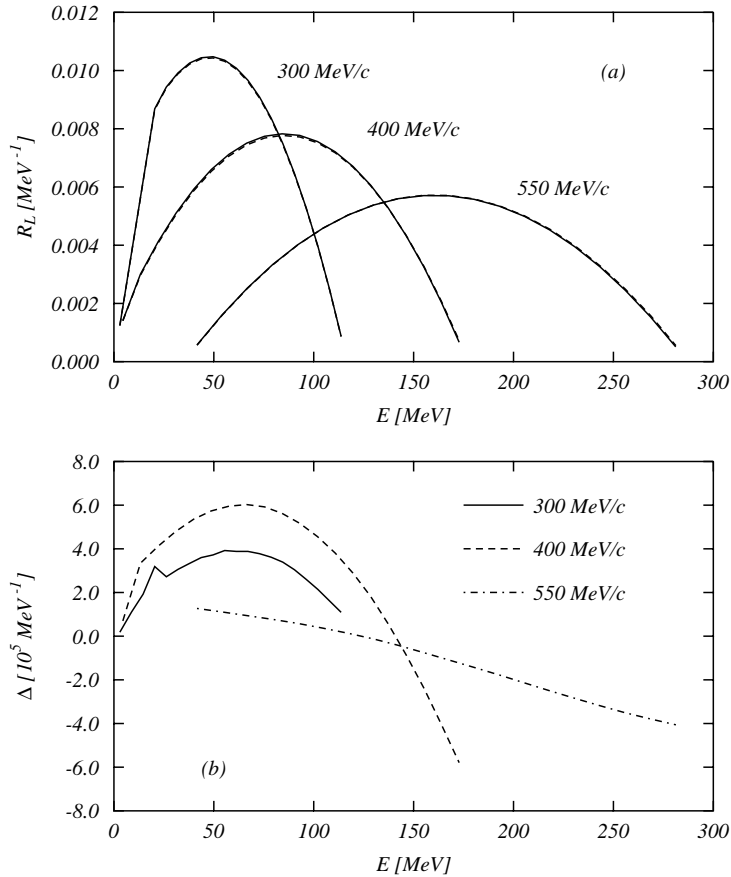}
\end{center}
\vspace{1.5cm}
\caption{\small In $(a)$ the nuclear matter proton responses
  calculated with our model (dashed lines) are compared with those
  obtained with a full FHNC calculation (full lines). The lower panel
  show the difference between the responses calculated in the two
  approaches.  }
\label{comp}
\end{figure}

We have compared the nuclear matter charge responses
calculated with our model \cite{ama98} with those 
obtained with the FHNC procedure. The two calculations have been done
with  the gaussian correlation
function of $^{12}$C of fig. \ref{corr1}.
The comparison between the results of the two calculations
is shown in fig. \ref{comp}, where the dashed lines (our model) 
are almost exactly overlapped to the
full ones (FHNC).

The excellent agreement between the two calculations has been obtained
because we have included in our model, in
order to have a correctly normalized many-body wave function,
both two and three body terms. 
The effect of the three point diagrams can be seen in the figs. \ref{rl}
and 7 where longitudinal an transverse responses for infinite
nuclear matter and for the $^{12}$C nucleus are compared.
We have used also for these calculations  the
$^{12}$C gaussian correlation of fig. \ref{corr1}. The finite system
responses have been evaluated with the single particle wave
functions producing the density shown in fig. \ref{dens}.

In fig. \ref{rl}, the left panels show the nuclear matter longitudinal
responses for three different values of the momentum transfer. 
The full lines are the uncorrelated responses, i.e. the Fermi
gas responses. The dotted lines have been obtained by evaluating only the
two-points diagrams, while the results of the calculations where both
two and three points diagrams have been taken into account are shown
by the dashed lines. The effect of the inclusion of the three point
diagrams has opposite sign with respect to that of the two points
diagrams. This fact is not surprising since, in the normalization
of the charge, the contribution of the the three points diagrams
cancels exactly that of the two points diagrams.
The finite size effect do not modify this conclusion as one can
observe in the three right panels of the figure.

The calculation of the transverse response has been done by
considering only the one-body magnetization current. The left panels
of fig. 7 show the nuclear matter responses. Also in this case
the effect of the three--points diagrams has opposite sign with
respect to that of the two-points diagrams. We observe that the final
response is extremely close to the uncorrelated response. The
different behavior with respect to the case of the
longitudinal response is produced
by the fact that, for the nuclear matter transverse response, the 
diagrams $(d)$ and $(l)$ of fig. \ref{diag2} do not contribute. 
This is strictly true only for the infinite system. 
In finite systems, only the diagram $d$ is zero.
The right panels of
fig. 7 show that this is the main responsible for the different
behavior of the correlations in the longitudinal and the transverse
response.  

The model has been extended also to the case when the final state has
two particles in the continuum \cite{co98},
by using the same procedure described above for
the 1p-1h excitation. 
In fig. \ref{rl2ph} we show the longitudinal 2p-2h responses for three
different values of the momentum transfer . 
The dashed lines have been obtained considering only the two
points diagrams, while the
full lines show the results obtained considering both two and three
points diagrams. Also in this case the contribution of the
three-points diagrams is relevant, close to the 40\% for some value of
the excitation energy.

It is necessary to remark that the absolute value of the response is
three order of magnitude smaller than the 1p-1h response.

%
%
\begin{figure}[ht]
\begin{center}
\vspace*{2.5cm}
\hspace*{-1.5cm}
\leavevmode
\epsfysize = 300pt
\epsfbox[70 200 500 650]{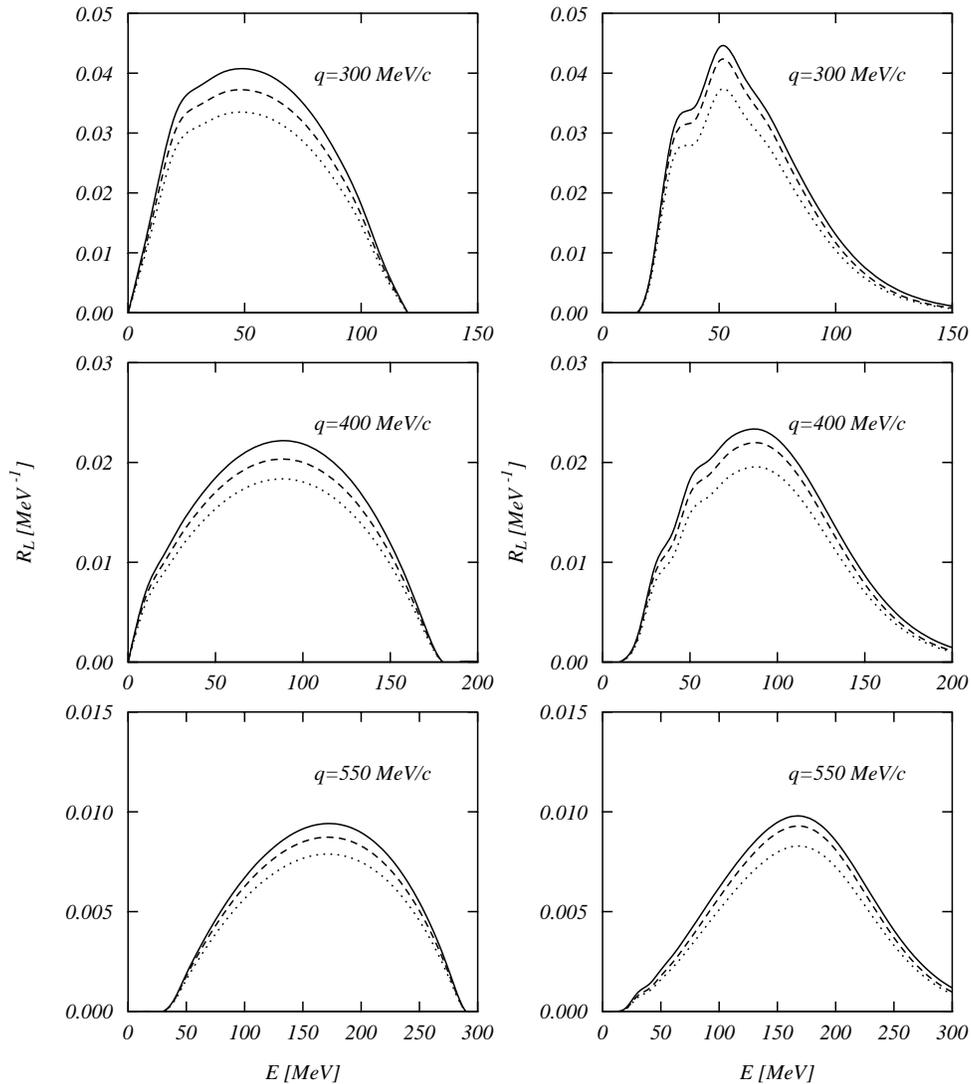}
\end{center}
\vspace{2.5cm}
\caption{\small Longitudinal responses at various values of the momentum
  transfer calculated for 1p-1h excitation.
  The full lines represent the uncorrelated response the dotted lines
  the responses obtained by adding the two-points diagrams, and
  the dashed lined those  obtained including also the three--points
  diagrams. The curves shown in the three left panels have been
  obtained for an infinite system while those shown in the right
  panels have been obtained for $^{12}$C using the wave functions
  giving the densities of fig. \protect\ref{dens}.}
\label{rl}
\end{figure}

%
%
\begin{figure}[ht]
\begin{center}
\vspace*{1.8cm}
\hspace*{-1.5cm}
\leavevmode
\epsfysize = 300pt
\epsfbox[70 200 500 650]{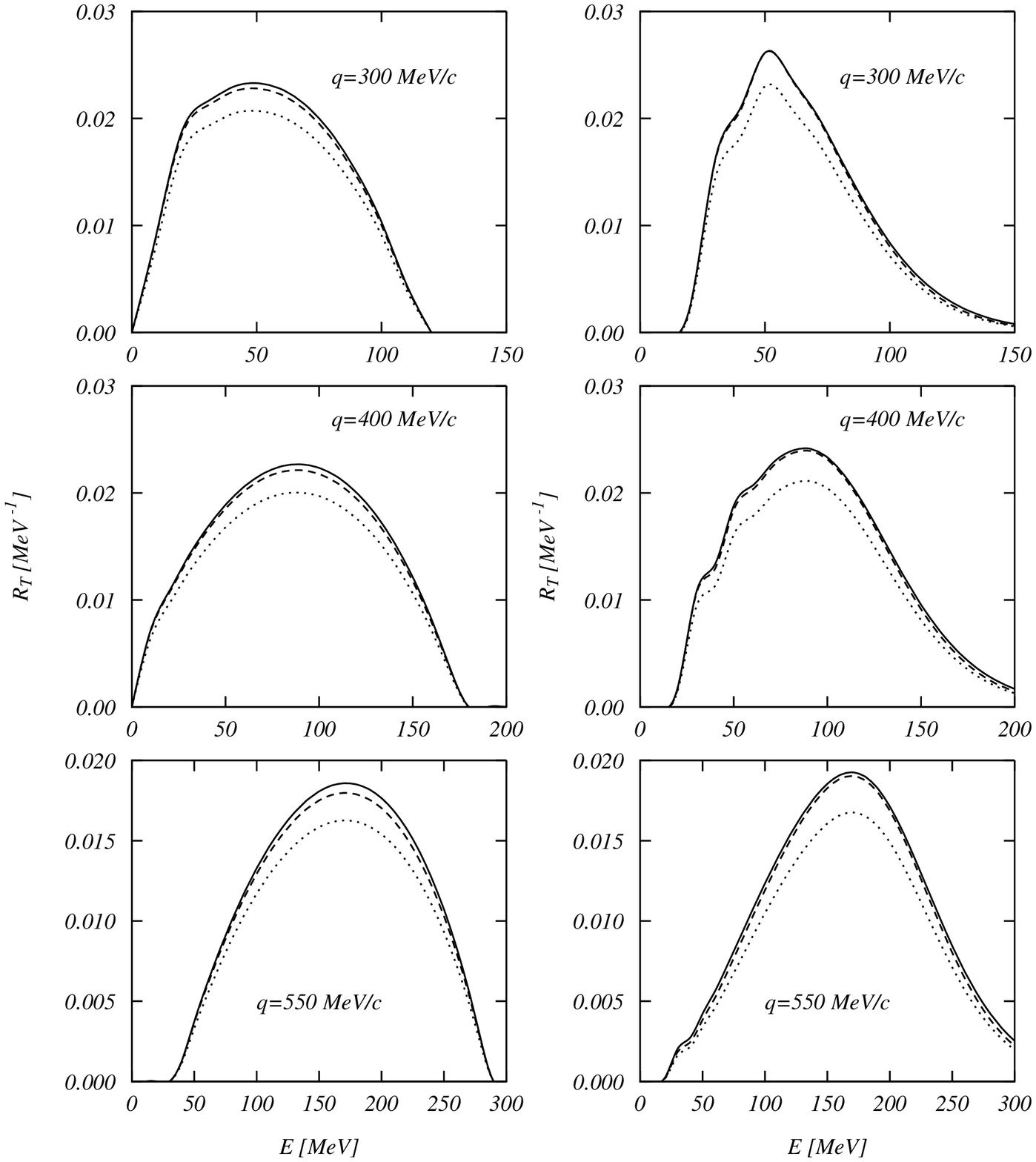}
\end{center}
\setcounter{figure}{7}
\label{rt}
\end{figure}
\vspace*{2.cm}
\centerline
{\small Figure 7. Same as in fig. \protect\ref{rl}
for the transverse responses. } 

\vspace{.5cm}

\section{Summary and conclusions}

We have developed a model to describe nuclear responses taking into
account short--range correlations. 
We have shown that mean values of the density operator can be
calculated with minimal error by using a cluster expansion truncated
up to the first order terms in the function $h(r_{ij})$. This
truncation should be properly done to maintain the correct
normalization of the wave function. This implies that both two and three
points diagrams should be considered.
We have shown that in both 1p-1h and 2p-2h responses the
quantitative contribution of the three-point diagrams is relevant
and cannot be simulated by multiplying the responses by a
constant renormalization factor.

The calculations we have presented are limited to inclusive
responses, but we plan to apply the model to treat 
exclusive processes.

\vskip 0.5 cm
\noindent{\bf Acknowledgments}\\[1ex]
This work has been partially supported by the agreement CICYT-INFN.
S.R. would like to thank the Italian Foreigner Ministery for the
financial support.

%
%
\begin{figure}
\begin{center}
\vspace*{2cm}
\hspace*{-2.0cm}
\leavevmode
\epsfysize = 250pt
\epsfbox[70 200 500 650]{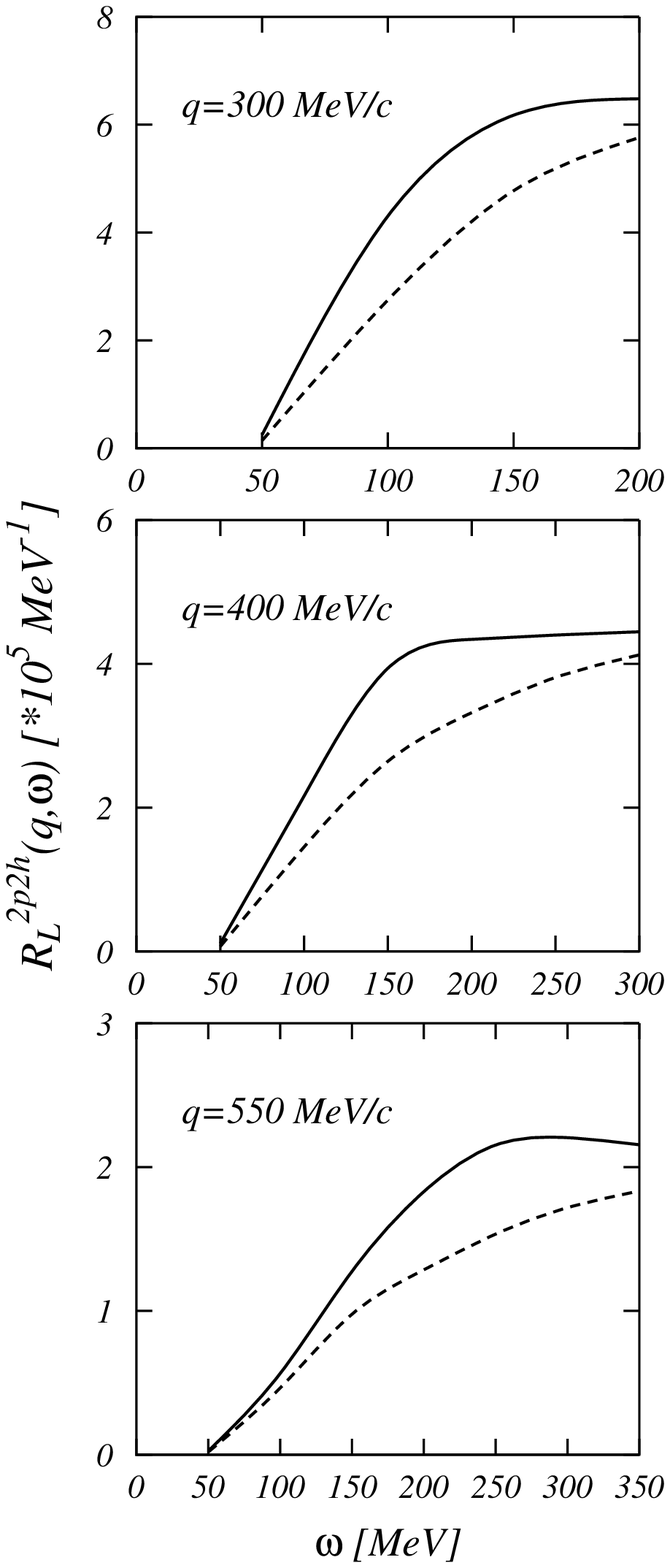}
\end{center}
\vspace{2cm}
\caption{ Charge response for 2p-2h excitation. The dashed lines
  represent the contribution of the two-points diagrams only, while
  the full lines show the result of the full calculation
}
\label{rl2ph}
\end{figure}

\end{document}